%% file: main.tex
\documentclass[sigconf]{acmart}
\input{macro}

\copyrightyear{2026}
\acmYear{2026}
\setcopyright{cc}
\setcctype{by}
\acmConference[SBFT '26]{19th Search-Based and Fuzz Testing Workshop }{April 12--18, 2026}{Rio de Janeiro, Brazil}
\acmBooktitle{19th Search-Based and Fuzz Testing Workshop (SBFT '26), April 12--18, 2026, Rio de Janeiro, Brazil}
\acmPrice{}
\acmDOI{10.1145/3786155.3795701}
\acmISBN{979-8-4007-2387-2/2026/04}

\AtBeginDocument{}

\begin{document}

\title{AutoRestTest at the SBFT 2026 Tool Competition}

\author{Tyler Stennett}
\email{tyler.stennett@gatech.edu}
\affiliation{%
  \institution{Georgia Institute of Technology}
  \city{Atlanta}
  \state{Georgia}
  \country{USA}
}

\author{Myeongsoo Kim}
\email{mysoo@amazon.com}
\affiliation{%
  \institution{AWS AI Labs}
  \city{Santa Clara}
  \state{California}
  \country{USA}
}

\author{Saurabh Sinha}
\email{sinhas@us.ibm.com}
\affiliation{%
  \institution{IBM Research}
  \city{Yorktown Heights}
  \state{New York}
  \country{USA}
}

\author{Alessandro Orso}
\email{orso@uga.edu}
\affiliation{%
  \institution{University of Georgia}
  \city{Athens}
  \state{Georgia}
  \country{USA}
}

\renewcommand{\shortauthors}{Stennett et al.}

\begin{abstract}
Large input spaces and complex inter-operation dependencies make black-box REST API testing challenging. AutoRestTest combines a Semantic Property Dependency Graph, multi-agent reinforcement learning, and large language models to intelligently explore large API input spaces. In the SBFT 2026 REST League, AutoRestTest ranked first in all three evaluation categories---fault detection, overall efficiency, and overall effectiveness---on 11 APIs (317 operations, approximately 29 per API), averaging 67.09 unique server errors and 17.27 successfully processed operations per API under a one-hour testing budget.
\end{abstract}

\keywords{Automated REST API Testing, Multi-Agent Reinforcement Learning, Large Language Models}

\maketitle

\section{Introduction}
REST APIs are central to modern web services, making automated testing of these APIs a critical research area. Most of the tools in this space implement a black-box testing approach: they use the OpenAPI Specification (OAS)~\cite{openapi-3.2.0} to exercise endpoints without source code access. To evaluate and distinguish leading tools, the REST League competition~\cite{restleague} at SBFT 2026 provides a standardized benchmark featuring diverse services and robust evaluation criteria.

Competing against five other state-of-the-art black-box REST API testing tools, \name achieved first place across all three evaluation categories---fault detection, overall efficiency, and overall effectiveness. Its success stems from three key components: a Semantic Property Dependency Graph (SPDG) that captures inter-operation dependencies, a multi-agent reinforcement learning (MARL) system with four specialized agents, and a large language model (LLM)-driven value generation module that produces domain-aware inputs. \name is publicly available~\cite{artifact}.

\section{\name}
\label{sec:approach}

\begin{figure*}[t]
  \centering
  \includegraphics[width=0.72\textwidth]{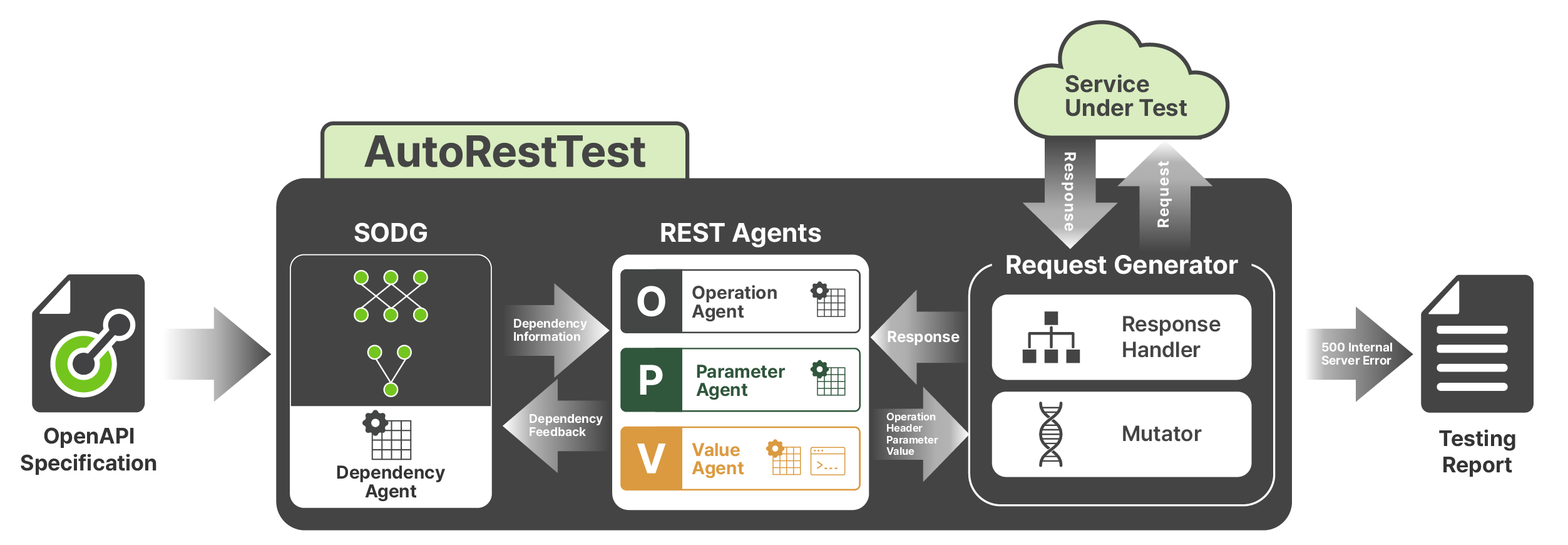}
  \vspace{-7pt}
  \caption{Overview of \name~\cite{kim2025multi}.}
  \vspace{-9pt}
  \label{fig:architecture}
\end{figure*}

\name's workflow is depicted in Figure~\ref{fig:architecture}. Since its initial presentation~\cite{stennett2025autoresttest, kim2025multi}, \name has been refined in several ways. LLM preprocessing is now parallelized across operations, reducing setup time. The parameter agent explicitly includes all required parameters and applies stratified sampling over optional parameters for balanced coverage. New boundary value mutations target edge cases (e.g., empty strings, extreme numeric values) to trigger server errors. Finally, broader LLM-provider compatibility supports cheaper and open-source models, and improved OAS parsing increases robustness across diverse specifications.

\vspace{-0.6em}
\subsection{Semantic Property Dependency Graph}
\name parses the OAS and constructs a graph with each API operation as a vertex. Using a lightweight GloVe embedding model~\cite{pennington2014glove}, it measures semantic similarity between parameters, request bodies, and responses across operation pairs, adding edges where similarity exceeds a configurable threshold. This weighted graph captures parameter dependency strength and is updated dynamically during testing as new dependencies are discovered.

\vspace{-0.6em}
\subsection{REST Agents}
\name employs four specialized agents: the \textit{operation agent} selects the next endpoint to explore, the \textit{parameter agent} uses stratified sampling to choose which parameters to include, the \textit{value agent} determines data sources and assigns values (from LLM-generated candidates, previously successful responses, or defaults), and the \textit{dependency agent} maintains the SPDG and tracks inter-operation dependencies. While \name also supports a header agent for basic bearer token authentication, it was disabled for the contest.

Each agent maintains a Q-table and selects actions via an epsilon-greedy strategy with a configurable learning rate and discount factor.

\vspace{-0.6em}
\subsection{LLM-Driven Value Generation}
To minimize token usage during testing, \name performs LLM-based value generation in a parallelized preprocessing phase. For each operation, the tool queries the LLM to produce domain-aware parameter values and validates them against the System Under Test (SUT). If the validation fails, server responses inform a refined query; this process repeats for up to three attempts per operation. The successfully validated values---or, if all attempts fail, the values from the final retry---are stored in the value agent's action space for efficient reuse across test generation.

\vspace{-0.6em}
\subsection{Request Generation}
With agents initialized and values preprocessed, \name begins constructing requests. Each agent independently selects its action, and the resulting request is dispatched to the SUT. The response status code drives a shared reward signal: 2xx responses are prioritized to build successful request chains, while 5xx responses are rewarded to encourage fault detection. All other responses are penalized to steer agents toward productive exploration of the API state space, and Q-tables are updated accordingly.

Finally, to uncover edge cases and internal server errors, \name applies mutations to refined request payloads. Mutations include parameter type alterations, name modifications, media type changes, and boundary value substitutions.

\section{Evaluation in the REST League}
We submitted \name to the REST League~\cite{restleague} competition at SBFT 2026. The evaluation benchmark of the competition consists of 11 real-world REST APIs with 317 operations (approximately 29 per API). The competition organizers executed each tool 10 times per API with a one-hour time budget, averaging the results to account for non-determinism.

\vspace{-0.25em}
\paragraph{Configuration}
We configured \name with a semantic similarity threshold of 0.7, epsilon-greedy exploration starting at $\epsilon=1$ and decaying to 0.2, learning rate $\alpha=0.1$, discount factor $\gamma=0.9$, and mutation probability of 0.2. For LLM-based value generation, we set the temperature to 1 to encourage diverse outputs, and used the low-cost Gemini 2.5 Flash Lite model~\cite{gemini-2-5-flash-lite} to manage expenses across runs in the absence of caching.

\vspace{-0.25em}
\paragraph{Evaluation Criteria}
REST League defines three Z-score-normalized evaluation categories, termed challenges~\cite{restleague}: \textit{fault detection} (unique 5xx errors), \textit{overall efficiency} (AUC over time for faults, operations, and coverage), and \textit{overall effectiveness} (combined end-of-execution fault detection, operation coverage, and branch coverage).

\vspace{-0.25em}
\paragraph{Results}
\name achieved the highest score in all three categories against CATS, EvoMaster, RESTest, RestTestGen, and Schemathesis. Most notably, it detected 2.7$\times$ more faults on average (67.09) than the next-best tool in that category (24.65), and achieved a 2.3$\times$ higher fault detection AUC (151,490) than the next-closest tool (66,761), indicating that its MARL architecture and epsilon-decay strategy produce effective tests quickly---critical for time-constrained environments. Beyond fault detection, \name also covered the most operations on average (17.27), demonstrating breadth without sacrificing depth. Full results are detailed in the competition report~\cite{restleague}.

Across all runs, \name consumed approximately 9.6M input tokens and 2.2M output tokens, totaling about \$1.83, or roughly \$0.02 per run per service.

Despite strong overall performance, \name encountered a crash on the \texttt{\small flight-search} API due to an implicit UTF-8 encoding assumption in its response parser. Future versions will handle diverse response encodings more gracefully.

\section{Conclusion}
We presented \name and its performance at the SBFT 2026 REST League competition~\cite{restleague}, where it ranked first in fault detection, overall efficiency, and overall effectiveness by combining dependency-aware exploration via the SPDG, adaptive test generation through MARL, and LLM-driven value generation---finding 2.7$\times$ more faults than the next-best tool at roughly \$0.02 per run per service. Future work will target more resilient response parsing and richer reward signals to improve robustness across diverse APIs. \name is open source and publicly available~\cite{artifact}.

\bibliographystyle{ACM-Reference-Format}
\bibliography{references}

\end{document}

%% file: macro.tex
\usepackage{xspace}

\newcommand{\name}{\textsc{AutoRestTest}\xspace}